\documentclass{aastex631}

\usepackage{amsmath}
\usepackage{graphicx}
\usepackage{multirow}

\shorttitle{Anisotropic Diffusion Model for Pulsar Halos}
\shortauthors{Yan et al.}

\begin{document}

\title{Anisotropic Diffusion of $e^\pm$ in Pulsar Halos over Multiple Coherence of Magnetic Fields}

\author{Kai Yan}
\affiliation{School of Astronomy and Space Science, Nanjing University, Nanjing 210023, China}
\affiliation{Key laboratory of Modern Astronomy and Astrophysics (Nanjing University), Ministry of Education, Nanjing 210023, China}

\author{Sha Wu}
\affiliation{Key Laboratory of Particle Astrophysics \& Experimental Physics Division \& Computing Center, Institute of High Energy Physics, Chinese Academy of Sciences, 100049 Beijing, China}
\affiliation{Tianfu Cosmic Ray Research Center,
 Chengdu 610000, Sichuan, China}

\author{Ruo-Yu Liu}
\affiliation{School of Astronomy and Space Science, Nanjing University, Nanjing 210023, China}
\affiliation{Key laboratory of Modern Astronomy and Astrophysics (Nanjing University), Ministry of Education, Nanjing 210023, China}
\affiliation{Tianfu Cosmic Ray Research Center,
 Chengdu 610000, Sichuan, China}
 
\correspondingauthor{Ruo-Yu Liu}
\email{ryliu@nju.edu.cn}

\begin{abstract}
%The slow particle diffusion coefficient in pulsar halos, as inferred from the observed TeV gamma-ray intensity profiles, is attributed to the cross-field diffusion in the anisotropic diffusion model, assuming the interstellar turbulence is sub-Alfvénic in the surrounding ISM and the line-of-sight (LOS) to the pulsar is roughly aligned with the local mean magnetic field direction in the halo. 
The slow particle diffusion in pulsar halos, inferred from TeV gamma-ray surface brightness profiles, is attributed to cross-field diffusion under the anisotropic diffusion model. This model assumes sub-Alfvénic interstellar turbulence in the surrounding medium of the pulsar and a rough alignment of the line-of-sight of observers towards the pulsars with the local mean magnetic field direction in the halo. In this model, the expected morphology of a pulsar halo is highly dependent on the properties of the interstellar magnetic field. We investigate the anisotropic diffusion of electron-positron pairs across multiple coherence of magnetic fields in pulsar halos in this work. We focus particularly on their influences on the predicted gamma-ray surface brightness profile and the asymmetry of the halo's morphology, as well as the observational expectations by the Large High Altitude Air Shower Observatory (LHAASO). Our results indicate that the requirement of a specific magnetic field geometry can be alleviated when accounting for a limited (and realistic) coherence length of the magnetic field in the model. Also, the halo's morphology may appear less asymmetric, especially after being smoothed by the point spread function of instruments. It largely relaxes the tension between the asymmetric morphology of halos predicted by the model and lack of apparent asymmetric halos detected so far. Our findings demonstrate the important influence of the coherence length of interstellar magnetic field on the distribution of particles around their accelerators, and the consequence on the measured source morphology.
\end{abstract}

%\keywords{TeV halos; anisotropic diffusion; LHAASO observation}

\section{Introduction}
Pulsar halos are the relativistic pair-rich region around middle-aged pulsars, where pulsars no longer dominate the dynamics \citep{2020A&A...636A.113G, Liu22_review}. These halos result from electrons and positrons (hereafter we do not distinguish positrons from electrons for simplicity), which have escaped from the pulsar wind nebulae into the ambient interstellar medium (ISM), up-scattering the cosmic microwave background (CMB) and the interstellar infrared radiation field \citep{2017PhRvD..96j3016L, Lopez22NA, Fang22}. To fit the observed intensity profile with a simple isotropic diffusion model, an extremely small diffusion coefficient of $D \lesssim 10^{28} \, \rm cm^2 s^{-1}$ (at an electron energy of $\sim 100 \, \rm TeV$) is needed \citep{abeysekara17extended,LHAASO0622}. This diffusion coefficient is significantly smaller than that inferred from the measurements on the local secondary-to-primary cosmic-ray (CR) ratios by about two orders of magnitude \citep[e.g.][]{2010ApJ...722L..58S}. X-ray observations suggest the magnetic field in pulsar halos may be weaker than the typical ISM value \citep{2019ApJ...875..149L, Khokhriakova23, Manconi24}, leading to the requirement of an even smaller diffusion coefficient.

The origin of such a slow diffusion zone is still unclear. \citet{Lopez18} suggested that an external turbulence of injection scale $\lesssim 1$\,pc may produce the slow diffusion coefficient, which is much smaller than the typical injection scale of the interstellar turbulence. Other hypotheses propose amplification of turbulence either by  relativistic electrons themselves responsible for halo's emission via resonant \citep{2018Evoli,2021arXiv211101143M} or non-resonant \citep{Schroer22} streaming instability, or by related supernova remnants  \citep{2019MNRAS.488.4074F, 2021arXiv211101143M}. So far, a common consensus on how the turbulence is amplified has not been reached. 

Alternatively, \citet{Liu19prl} proposed an anisotropic diffusion model assuming the turbulence of the ambient ISM to be sub-Alfv{\'e}nic with the Alfv{\'e}nic Mach number $M_A$ less than unity. In the sub-Alfv{\'e}nic regime, the magnetohydrodynamic (MHD) turbulence is expected to be dominated by a relatively strong mean magnetic field. In the anisotropic diffusion model, the electron diffusion perpendicular to the mean magnetic field can be much slower than that along the mean magnetic field, resulting in a cylindrical symmetry of the electron density with respect to the magnetic field direction. Given such a spatial distribution, the viewing angle $\phi$ (the angle between the observer's LOS towards the pulsar and the mean magnetic field direction) is crucial to the apparent morphology of the halo. While the anisotropic diffusion model can naturally explain the slow diffusion and the non-detection of X-ray emission \citep{Liu19_Xray, Manconi24} from Geminga’s pulsar halo, it requires a relatively small viewing angle (approximately less than $5^\circ$) to reproduce the observations. If the mean magnetic field directions around these pulsars are randomly oriented, the probability of finding a pulsar halo with a small viewing angle would be rare. Therefore, most TeV halos are supposed to show asymmetric or elongated morphology under the anisotropic diffusion model, while most of pulsar halos or candidate sources detected so far are not reported with clear asymmetric morphology \citep{Liu19prl, DeLaTorreLuque22, KaiYan22apj}. 

Given such a tension, \citet{KaiYan22apj} suggest that pulsar halos with smaller viewing angles are more easily detected as they appear more compact and bright, resulting in a selection effect favoring spherically symmetric halos or those with small viewing angles. Apart from the selection effect, taking into account a limited coherence length of the interstellar magnetic field might also explain the lack of apparent asymmetric pulsar halos in observations. Indeed, the mean interstellar magnetic field cannot always keep the same direction. In general, the coherence length of interstellar magnetic field $l_{\rm c}$ ranges in $\sim 10-200 \, \rm pc$ \citep{2009Cho_lc, 2010Chepurnov_lc, 2016Beck_lc}. If the coherence length is not sufficiently long within a pulsar halo, particles are expected to traverse multiple magnetic coherence lengths, among which the preferential direction of diffusion changes significantly. As a result, the particle distribution within the pulsar halo may become more isotropic. Most previous studies considered particle diffusion within only one coherence length (or equivalently assuming $l_{\rm c}=+\infty$) \citep{Liu19prl,KaiYan22apj,QizuoWu23apj}. While this approach simplifies the modeling process, it may overlook significant features that arise from the propagation of particles in multiple coherence lengths. The importance of considering multiple coherence lengths in magnetic fields has recently been studied by \citet{Fang23prd} and \citet{Bao24b,YiweiBao24}. \citet{Fang23prd} consider a simplified geometry of the magnetic field and indicated the presence of a ``wing'' component in the resulting halo morphology when considering diffusion over multiple coherence. \citet{Bao24b,YiweiBao24} proposed that particles propagating across multiple coherence could result in a few hotspots, or the so-called ``mirage sources'', around the particle accelerator, which may result from the projection effect of particles distributing along the field lines with small viewing angle to the observers's LOS.

In this work, we employ the random walk simulation to simulate the particle diffusion across multiple magnetic coherence. This method allows us to model the stochastic nature of particle propagation and account for the varying diffusion properties of particles over different magnetic coherence. By doing so, we aim to provide a more comprehensive understanding of the particle propagation and distribution in pulsar halos, as well as the consequent morphology of pulsar halos under the anisotropic diffusion model. Additionally, we carefully consider LHAASO's instrumental responses and simulate the expected significance maps. These simulations enable us to make direct comparisons with observational data. By systematically exploring the effect of coherence lengths on observational characteristics, we aim to investigate the impact of magnetic field structures on the observable features of pulsar halos and cosmic ray propagation.

The rest of the paper is organized as follows: in Section 2, we introduce the method to obtain the particle distribution from random walk simulation and evaluate the detection capacity of LHAASO on these sources; in Section 3, we present and discuss the main results of the paper; we further discuss some uncertainties of model parameters in Section 4 and give our conclusion in Section 5.

\begin{figure*}[htbp]
\hspace{-0.5cm} \includegraphics[width=1\textwidth]{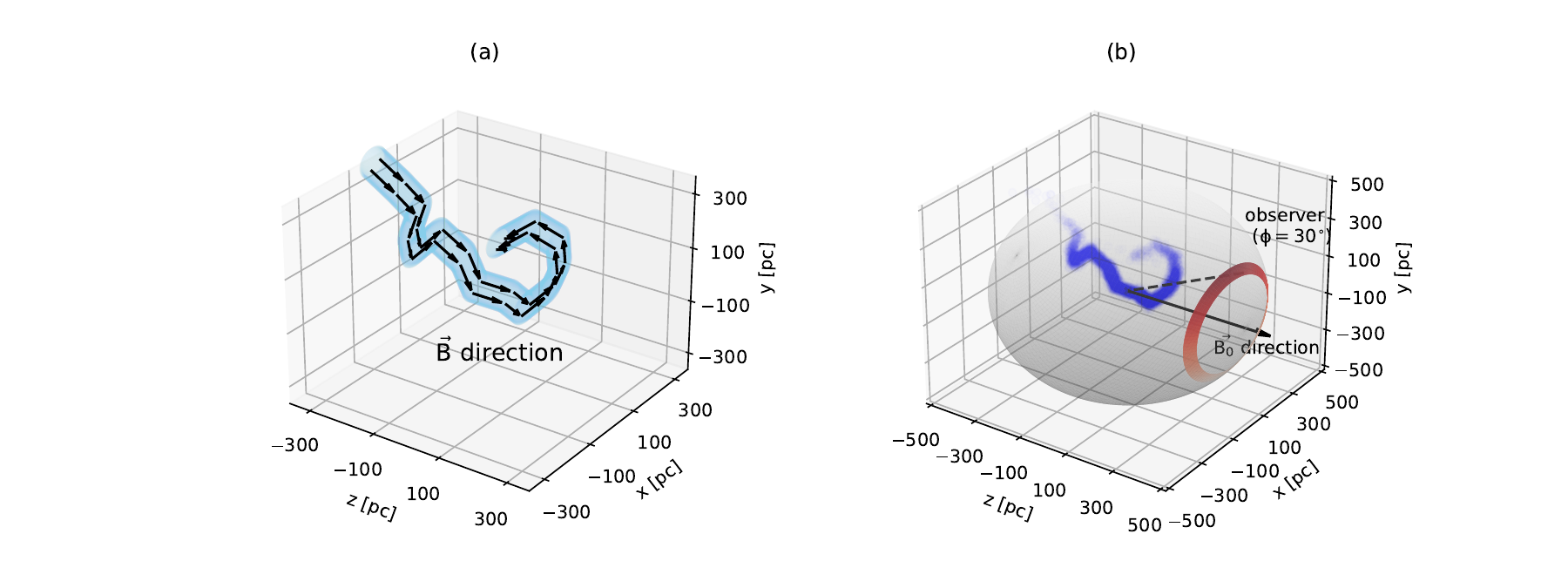}
\caption{(a) A schematic figure of the configuration of the multi-coherence of the mean magnetic field.  coherence length is set to be 100\,pc in this figure. The light blue regions show the coherent magnetic regions within which the mean magnetic field direction are indicated by the black arrows. (b) Particle distribution in the magnetic field. $B_0$ represents the central coherence of the magnetic field. The red band illustrates the observer with a viewing angle of $\rm 30^{\circ}$ with respect to the direction of $B_0$.}
\label{fig:B_config}
\end{figure*}

\section{Methods} \label{sec:style}

\subsection{Particle Transport and Radiation}
The transport of escaping electrons in a pulsar halo is dominated by diffusion. The transport equation can be given by
\begin{equation}
\frac{\partial N(\vec{r}, E_e, t )}{\partial t}=\nabla \cdot[\mathbb{D}(\vec{r}, E_e) \cdot \nabla N(\vec{r}, E_e, t )]
-\frac{\partial}{\partial E_{\rm e}}\left(\dot{E}_{\rm e} N(\vec{r}, E_e, t )\right)+Q\left(E_{\rm e}\right) s(t) \delta(\vec{r})
\label{E1}
\end{equation}
where $N(\vec{r},E_e,t)$ is the particle number density, $\mathbb{D}(\vec{r},E_e)$ is the diffusion tensor. $\dot{E}_{\rm e}$ is the cooling rate of electrons through the inverse Compton (IC) radiation in the CMB and the interstellar infrared radiation field, and through the synchrontron radiation in the interstellar magnetic field. The last term in the equation represents the injection term from the pulsar wind nebula which is approximated as a point source.  The temporal behavior of the injection rate $s(t)$ is denoted as $s(t)=(1+t_{\rm age}/\tau)^2(1+t/\tau)^{-2}$ with $\tau$ being the spin-down timescale of the pulsar and $t_{\rm age}$ is the age of the pulsar. $\delta(\vec{r})$ is the Dirac function denoting the injection location. $Q(E_{\rm e})$ is the injection spectrum of electrons, which is assumed to be follow the form $Q(E_{\rm e})=N_{\rm 0}E_{\rm e}^{-p}e^{-E_{\rm e}/E_{\rm c}}$, with $p$ the spectral index and $E_{\rm c}$ the cutoff energy in the spectrum. We set $p=1.6$ and $E_{\rm c}=200 \, \rm TeV$ in this study. 

The injection normalization constant $N_{0}$ can be determined by $\int_{E_{\rm min}}^{E_{\rm max}} E_{\rm e} Q(E_{\rm e})d E_{\rm e} = \eta L_s$. $\eta$ is an input parameter called pair conversion efficiency, which represents the ratio of pulsar spindown energy that goes into pairs. $L_{\rm s}$ is the pulsar's current spindown power. We set the minimum energy $E_{\rm min}=0.1\,$TeV in the injection spectrum, noting that this value is not critical because most energy of electrons concentrate at high-energy end given $p=1.6$ and we mainly focus on $>10\,$TeV electrons which radiate in the observational energy range of LHAASO-KM2A. The maximum energy in the injection spectrum is set to $E_{\rm max}=1 \, \rm PeV$, which is also not crucial because the injection spectrum already features a cut-off at the energy $E_{\rm c}$.

The energy loss rate of electrons through a combination of synchrotron radiation and IC radiation can be given by
\begin{eqnarray}
\dot{E}_{\rm e}\equiv \frac{\mathrm{d} E_{\mathrm{e}}}{\mathrm{d} t}=-\frac{4}{3} \sigma_{\mathrm{T}} c\left(\frac{E_{\mathrm{e}}}{m_{\mathrm{e}} c^{2}}\right)^{2}\left[U_{\rm B}+\sum_{i}f_{\rm KN}(E_{\rm e})U_{\mathrm{i}}\right]
\label{eq:dEe}
\end{eqnarray}
where $\sigma_{\rm T}$ is the Thompson cross section, and $f_{\rm KN}=\left[1+(2.82 E_{\rm e}\epsilon_{\rm i}/m_{\mathrm{e}}^{2} c^{4})^{0.6}\right]^{-1.9/0.6}$ is a numerical factor accounting for the Klein-Nishina (KN) effect \citep{LHAASO21_nat}. The magnetic field strength is taken as 5$\, \rm \mu G$ and the magnetic energy density $U_{\rm \mathrm{B}}$ is $0.62\,\rm eV \ cm^{-3}$. $U_i$ and $\epsilon_i$ represent the energy density and the typical photon energy of the $i$th component of the interstellar radiation field (ISRF) assuming a black body or a grey body distribution with temperature $T_i$ for their spectra (i.e., $\epsilon_i=2.82kT_i$). The considered target radiation field includes the CMB radiation field ($T_{\rm CMB}$ = 2.73\,K and $U_{\rm CMB}$ = 0.26\,$\rm eV \ cm^{-3}$), a far-infrared radiation field ($T_{\rm FIR}$ = 30\,K and $U_{\rm FIR}$ = 0.3\,$\rm eV \ cm^{-3}$), and a visible light radiation field ($T_{\rm VIS}$ = 5000\,K, $U_{\rm VIS}$ = 0.3\,$\rm eV \ cm^{-3}$). The cooling timescale of electrons can be obtained by $t_{\rm c}=E_{\rm e}/\dot{E}_{\rm e}$.

\subsection{Diffusion Process with Random Walk Treatment}

We may obtain the electron distribution at the present time $t=t_{\rm age}$ by solving Eq.~(\ref{E1}), the solution of which reads
\begin{eqnarray}
N(\vec{r}, E_e)=\int_{\max \left[0, t_{\operatorname{tage}}-\tilde{t}_{\rm c}\left(E_0, E_e\right)\right]}^{\operatorname{tage}} \frac{\dot{E_e}\left(E_{0}\right)}{\dot{E_e}(E_e)} Q\left(E_{0}, t\right) \mathcal{H}(\vec{r}, E_e,t) \mathrm{d}t
\label{eq:E3}
\end{eqnarray}
where $\tilde{t}_{\rm c}\left(E_0, E_e\right)$ is the cooling time of an electron from the energy at injection $E_0$ to $E_{\rm e}$. Note that $E_0$ here is related to $E_e$ and $t$, and the relation can be found via Eq.~(\ref{eq:dEe}). $\mathcal{H}(\vec{r},E_e,t)$ describes the spatial distribution of the particles. The integration of $\mathcal{H}(\vec{r},E_e,t)$ over the whole space equals unity, which means that it is normalized to account for the total number of particles. This normalization ensures that the model accurately reflects the conservation of particle number across the spatial domain.

The form of $\mathcal{H}(\vec{r},E_e,t)$ is determined by the diffusion process. Analytical expressions for $\mathcal{H}(\vec{r},E_e,t)$ have been derived for some specific cases, such as one-zone or two-zone isotropic diffusion, and anisotropic diffusion within a single coherence length (see Appendix for details). However, for more complex scenarios involving spatially dependent diffusion, such as anisotropic diffusion across multiple coherence lengths, analytical solutions are not available. In this complex case, we employ the random walk simulation to numerically calculate the particle distribution. The random walk simulation is based on the idea that particles move in a series of random steps, with each step being independent of the previous one \citep{1928Einstein,1918vanderWaals}. In a random walk, a particle moves from its initial position to a new position in a random direction step by step, which is typically defined by a probability distribution. Over time, this series of random steps results in the diffusion of the particle.

With this treatment, the key to calculate the particle distribution in the context of multiple coherence of magnetic field is to obtain the space-dependent diffusion tensor.  
To achieve this, firstly we need to determine the mean direction of each magnetic coherence in the space. More specifically, we first generate a random direction of the mean magnetic field, where the particle accelerator is located at its center. The direction of the magnetic field maintains for a distance of $l_{\rm c}$. At each end of this segment of the magnetic field, we let the field direction shift a random inclination and the new direction also maintains for a length of $l_{\rm c}$. This process is repeated for many times until the total length of all generated magnetic coherence reaches kpc, which is longer than typical diffusion length (i.e, $2\sqrt{D_\parallel t_{\rm c}}$) of electrons with energy relevant for this study (above $10\,$TeV). In the end, we obtain something like a long magnetic flux tube composed of many segments of randomly oriented lines, which have the same length $l_{\rm c}$, connected end to end. 

In the left panel of Fig\,\ref{fig:B_config}, we show the configuration of the mean magnetic field in one realization of our simulation. According to \citet{HuirongYan08}, in the regime of sub-Alfv{\'e}nic MHD turbulence, the local perpendicular diffusion coefficient is given by $D_{\perp} = D_{\parallel}M_A^4$ within each coherence length, where $D_{\parallel }$ is the diffusion coefficient parallel to the mean magnetic field, $M_A\equiv \delta B /B$ is the Alf{\'e}nic Mach number, where $\delta B$ is the amplitude of the turbulent magnetic field at the injection scale, and $B$ is the strength of the mean (ordered) magnetic field. In this study, we set $D_{\parallel}=10^{28}(E_e / 1 \, \rm{GeV})^{1/3} \, \rm cm^2s^{-1}$ and $M_A=0.2$. A low $M_A$ value corresponds to a magnetically dominated environment, where the turbulence is weak and the magnetic field lines remain largely aligned with the mean magnetic field direction. Consequently, the coherence length $l_{\rm c}$ of the magnetic field in this case could be longer than that of a strong, isotropic turbulent flow, and is generally determined by the scale of the past astrophysical processes that shaping the environment where the halo resides (e.g., previous generations of supernova remnants, super bubbles, or other macroscopic astrophysical processes). In the Cartesian coordinate with $z-$axis aligned along the direction of the coherent magnetic field $B_0$ centred at the location of the injection point (i.e., the particle accelerator), the diffusion tensor $\mathbb{D}$ within each coherence can be obtained by multiplying the corresponding coordinate rotational matrix. 

We would like to caution that the magnetic field generated in this way is far from realistic. However, the main purpose of this work is to explore the effects of large-scale ($\gtrsim 50$\,pc) magnetic structures on particle diffusion. Therefore, instead of generating a realistic magnetic field from MHD simulations over a sufficiently large space, which would be extremely expensive, we focus on the large-scale magnetic field configuration that does not include small-scale turbulent features. It nevertheless allows us to study how the source morphology would look like if the preferential direction of particle diffusion changes several times within the typical diffusion length of particles. This is also the reason why we do not trace the trajectory of each electron in an MHD-based magnetic field following the Lorentz force as did in test-particle simulations, but approximate the diffusion process by random walks based on the space-dependent diffusion tensor. 

%\subsubsection{Model Setup and Comparison with Analytical Solutions}

After obtaining $N(E_e,\vec{r},t)$, we can calculate the IC emissivity of electrons $q_{\rm IC}(E,\vec{r})$ following the semi-analytical method given by \citet{Khangulyan14}. The 2D gamma-ray intensity profile, $I_\gamma (E,\theta, \xi)$, is the projection of the emission onto the plane-of-the-sky and can be obtained by integrating the emissivity over the observer's LOS towards each direction, as detailed in \citet{Liu19prl}.

\subsection{Estimation of LHAASO's detection}
To evaluate the morphology of a simulated pulsar halo detected by LHAASO, we need to estimate the signal-to-noise ratio (SNR) or the statistical significance of the halo's emission. Firstly, we need to convolve the theoretical 2D intensity profile with the point-spread-function (PSF) of LHAASO, which may blur the morphology of the halo observed by the instrument. A 2D Gaussian function is used to represent the PSF of LHAASO, so that the PSF-convolved 2D intensity profile can be given by
\begin{equation}
\begin{aligned}
\label{E5}
I_{\rm \gamma,\rm PSF} =  \iint \frac{1}{2\pi \sigma^{2}(E_\gamma)}\exp\left(-\frac{l^{'2}}{2\sigma^{2}(E_\gamma)}\right) I_{\rm \gamma }^{'}(E_{\rm \gamma},\theta^{'},\xi^{'})\sin\theta^{'}d\theta^{'}d\xi^{'} \end{aligned}
\end{equation}
where the angular distance $l^{'}=\arccos[\cos\theta \cos\theta^{'}+\sin\theta \sin\theta^{'}\cos(\xi -\xi^{'})]$ and $\sigma(E_\gamma)$ is the size of PSF as listed in \citet{KaiYan22apj}. The expected photon counts rate above $E_\gamma$ per solid angle towards the direction ($\theta$, $\xi$) from the pulsar's position can be given by 
\begin{equation}
C_\gamma(>E_\gamma, \theta, \xi)=\int_{E_\gamma}^\infty f_{\rm eff}(E_\gamma)I_{\gamma,\rm PSF}/E_\gamma^2dE_\gamma,
\end{equation}
where $f_{\rm eff}(E_{\gamma})$ represents the product of the effective area and the exposure time, calculated as 
\begin{equation}
f_{\rm eff}(E_{\gamma}) = T \int_{-t_{A50}}^{t_{A50}} A_{\rm eff}(\zeta,E_{\gamma})dt_A/24,    
\end{equation}
The integral evaluates effective area averaged over the change in zenith angle and weighted by the exposure time. In the integral, $\zeta$ is the zenith angle of a source (or a certain point in the sky), and it is related to the declination and the hour angle $t_{\rm A}$ of the source (or the certain point in the sky). $t_{\rm A50}$ in the lower and upper bounds of the integral means the hour angle when the zenith angle of the source is $50^\circ$. The bounds are set because events with $\zeta > 50^\circ$ are not used in LHAASO's analysis. One may find the relation among the hour angle $t_{\rm A}$, the declination $\delta$ and the zenith angle of the source $\zeta$ at LHAASO site by $\cos\zeta=\sin\delta \sin 29^\circ+\cos\delta \cos29^\circ \cos (15^\circ t_{\rm A}/{\rm hour})$, with $29^\circ$ being the declination of the LHAASO site. $A_{\rm eff}$ is the effective area of LHAASO-KM2A. We employ the effective area of the half-KM2A array shown in \citet{LHAASO21Chinesesci} and multiply it by a factor of 2 for $A_{\rm eff}$. $T$ is the operation time of LHAASO.

On the other hand, CR will contaminate the gamma-ray detection on the ground. At energy above 10\,TeV, the measurement of the muon component by LHAASO-KM2A provides an efficient rejection to the CR background and greatly improves the source detection capability. The simulated fraction of cosmic rays that pass the muon rejection cuts can reach as low as $\sim 0.01\%$ at several tens TeV and even below $0.001\%$ above 100\,TeV \citep{LHAASO19book}. The background counts rate above energy $E_\gamma$ per solid angle has been measured towards the direction of Crab Nebula \citep{LHAASO21crab}. We denote the background counts before muon-cut by $B_{\rm CR}$ and assume it homogeneous over the entire sky given the highly isotropic distribution of the CR arrival direction measured locally.

The numbers of on-source counts $N_{\rm on}$ and that of background counts $N_{\rm B}$ are calculated with
\begin{eqnarray}
&& N_{\rm on}=\mathcal{F}_{\rm poisson}\left( \lambda = \int_0^{\theta_{\rm s}}\int_0^{2\pi} (C_\gamma+B_{\rm CR}f_{cut})\sin\theta d\theta d\xi \right), \\
&& N_{\rm B}=\mathcal{F}_{\rm poisson}\left( \lambda = \int_0^{\theta_{\rm s}}\int_0^{2\pi} B_{\rm CR}\sin\theta d\theta d\xi \right)f_{cut}.  
\end{eqnarray}
Where $\mathcal{F}_{\rm poisson}(\lambda)$ denotes the Poisson distribution with an expectation of $\lambda$. $f_{\rm cut}$ is the simulated fraction of cosmic rays passing the hadron rejection cuts. 
%The coefficient 0.9 before the source's counts rate $C_\gamma(>E_\gamma, \theta,\xi)$ accounts for the fact that about 10\% gamma-ray events can not survive after applying for the background rejection cuts. 

We then estimate the significance of the pulsar halo according to the on-source and background counts. The significance of the source is calculated using a test statistic (TS) variable, defined as twice the logarithm of the likelihood ratios, i.e., ${\rm TS}=2\rm{ln}(\mathcal{L}_{\rm s+b}/\mathcal{L}_b)$, where $\mathcal{L}_{\rm s+b}$ is the maximum likelihood for the signal plus background hypothesis, and $\mathcal{L}_{\rm b}$ is the likelihood for the background-only hypothesis. According to the Wilks' theorem \citep{Wilks1938}, in the background-only case, the TS value follows a $\chi^2$ distribution of $n$ degrees of freedom, where $n$ is the number of free parameters in the signal model. For a point source with a fixed position, which has only one free parameter (the normalization when ignoring the spectral distribution), the pretrial significance is $\sqrt{\rm TS}$.

\section{Results} 
The main purpose of this work is to explore the effects of multi-coherence of magnetic field on the observational characteristics of pulsar halos. Therefore, we do not investigate the full parameter space of the magnetic field configurations, but focus on the influence of some key parameters such as coherence length $l_c$ and viewing angle $\phi$ on the observational properties of pulsar halos, such as the azimuthal-angle-averaged 1D surface brightness profile (SBP) and detectability of the asymmetry of the halo's morphology.

\begin{figure*}[htbp]
\includegraphics[width=\textwidth]{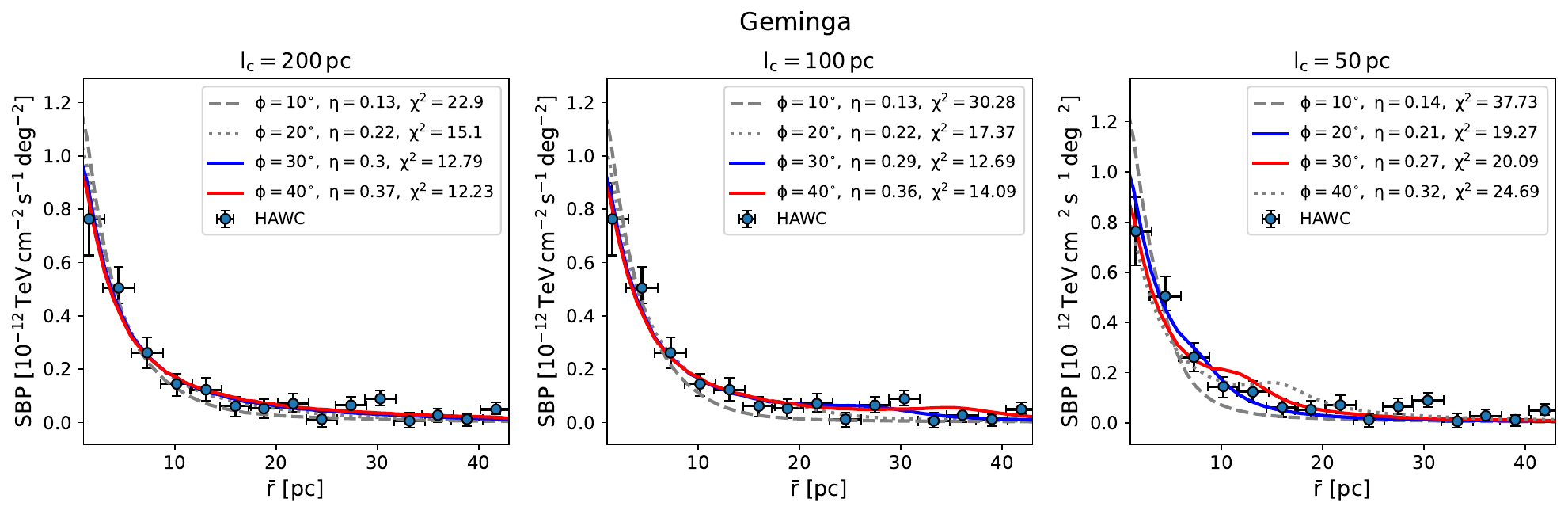}\\
\includegraphics[width=\textwidth]{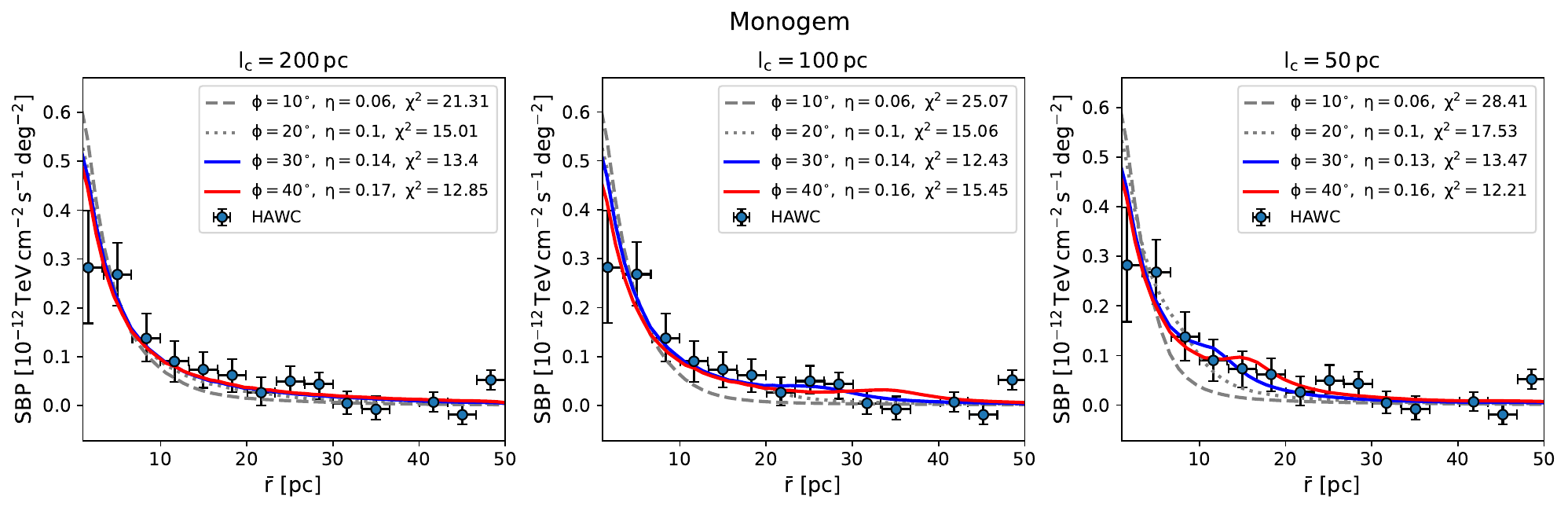}\\
\includegraphics[width=\textwidth]{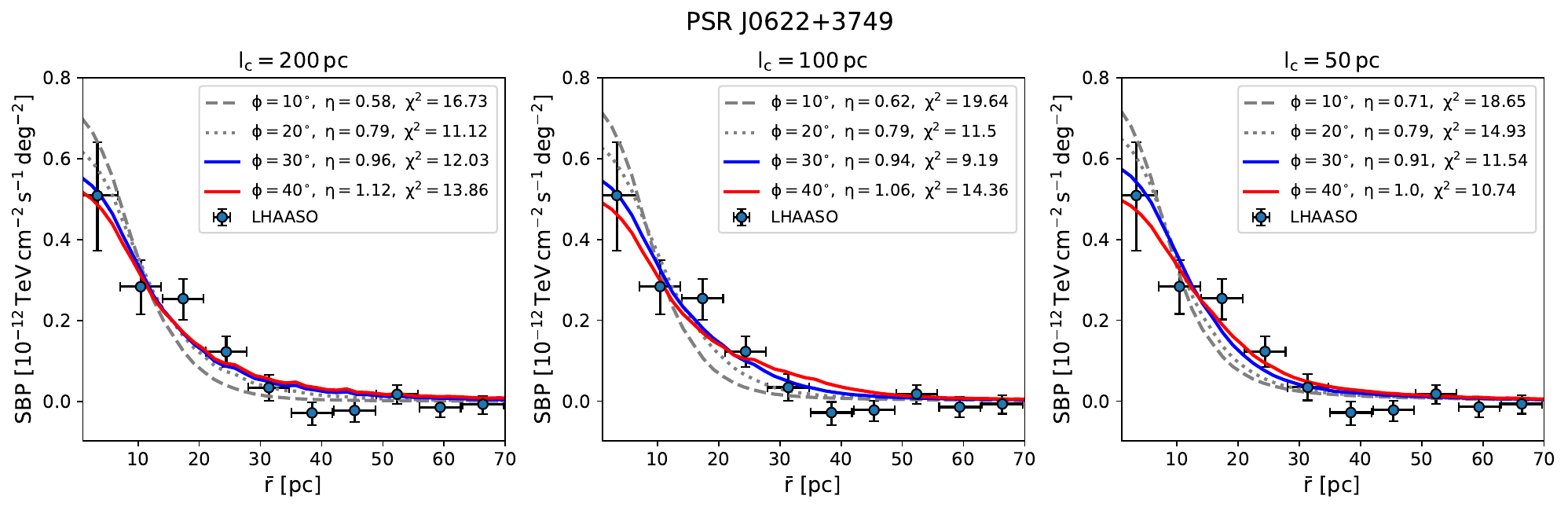}\\
\caption{SBPs of Geminga, Monogem, and PSR~J0622+3749. The coherence lengths decrease from 200 pc (left panels) to 100 pc (middle panels), and finally to 50 pc (right panels). The data points and error bars represent observations of HAWC on Geminga and Monogem in the top row and middle row respectively, and the observation of LHAASO on PSR~J0622+3749 in the bottom row. For Geminga and Monogem, we integrate over the spectrum in 8-40\,TeV, while for PSR~J0622+3749 we integrate over the spectrum above 25\,TeV, in order to match the observational energy range.}
\label{fig:SBP}
\end{figure*}

\begin{figure*}[htbp]
\includegraphics[width=0.33\textwidth]{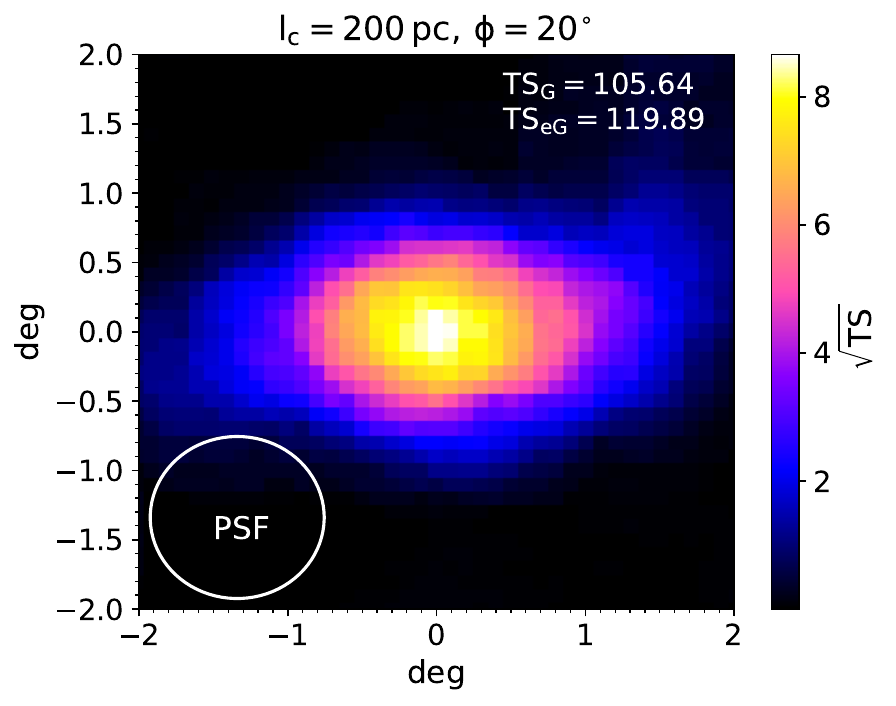}
\includegraphics[width=0.33\textwidth]{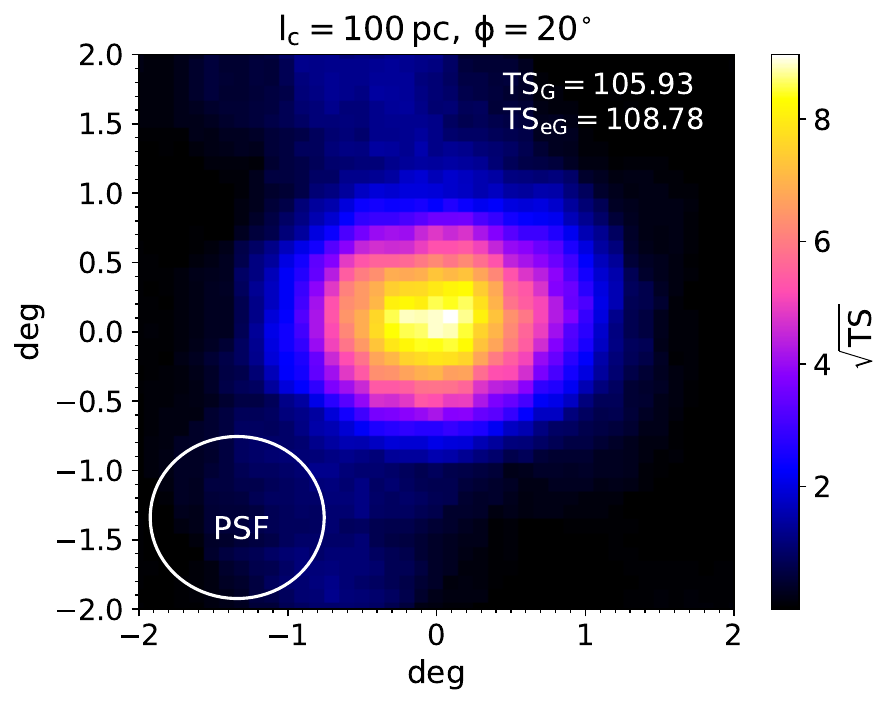}
\includegraphics[width=0.33\textwidth]{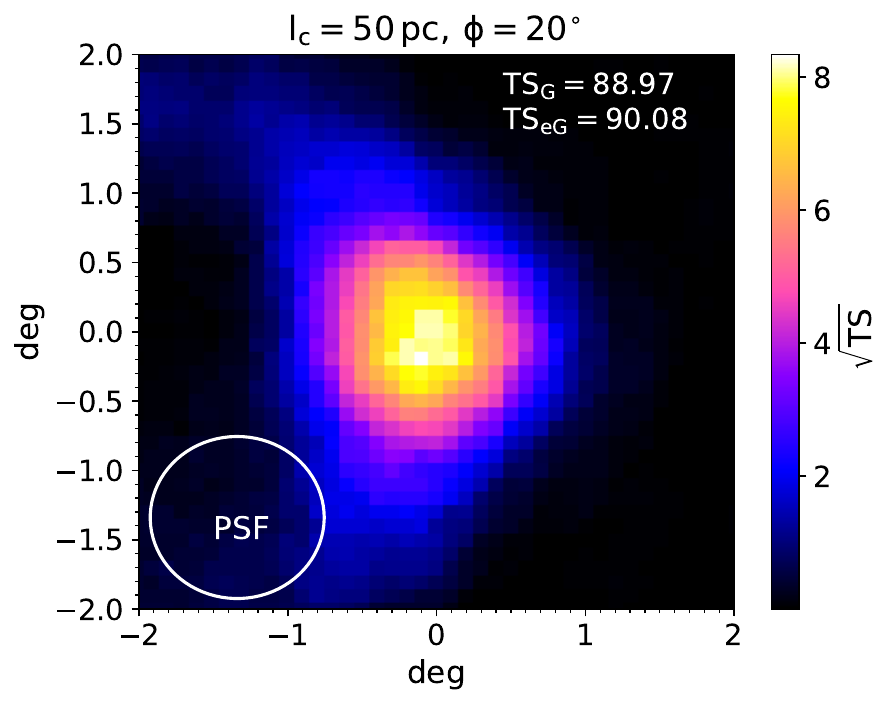}\\
\caption{Simulated significance maps of the $4^{\circ} \times 4^{\circ}$ regions around PSR J0622+3749 of LHAASO with energy above 25 TeV. The coherence lengths decrease from 200 pc (left panel) to 100 pc (middle panel), and finally to 50 pc (right panel). The white circle represents the 68\% containment radius of the LHAASO PSF. The upper corners note the best-fit test statistical ($TS$) values of the Gaussian template and the ellipse Gaussian template.}
\label{fig:0622morphology}
\end{figure*}

\subsection{Surface Brightness Profile}
In Fig.\,\ref{fig:SBP}, we present the fitting of the measured azimuthal-angle-averaged 1D SBPs for halos of Geminga, Monogem, and LHAASO~J0621+3755 under the anisotropic diffusion model, comparing results with different coherence lengths $l_c$. Note that the horizontal axis of each panel in this figure has been translated to the projected radius from the pulsar based on the distances of the pulsars (i.e, $d=250\,$pc for Geminga, 290\,pc for Monogem and 1.6\,kpc for PSR~J0622+3749). The analysis reveals significant differences in the SBP features as $l_c$ decreases. We observed that as $l_c$ decreases, the resulting SBPs exhibit two main changes\footnote{it also applies to the changes from $l_{\rm c}=+\infty$ to a limited $l_{\rm c}$}: 

%The transition from an infinite coherence length ($l_{\rm c} = +\infty$) to finite $l_c$ illustrates the impact of particle diffusion coherence scales on the observed profiles:

\begin{itemize}
    \item \textbf{SBPs becomes steeper at small angular radius while flatter at larger radius:} With a decreasing coherence length, particles propagated to the outer region of the halo are more likely guided by the magnetic field to other directions and do not concentrate on our LOS toward the vicinity of the pulsar. Their emission usually appears as  dimmer and diffusive wing-like structure at large angular distance, leading to a flatter SBP at large radius. The SBP at small radius then becomes dominated by radiation of electrons injected recently which are distributed close to the source center. Since the spatial distribution of particles near the injection point exhibits a larger gradient, the resulting SBP at smaller radius appears steeper (see also in Appendix Fig.\,\ref{fig:sp_phi}). From the perspective of SBP fitting, a steeper profile at small radius relaxes the constraint on the viewing angle (e.g. $\phi \le 5^{\circ}$ as discussed in \citealt{Liu19prl, 2022Luque}) and a much larger viewing angle can also match the observed data (see Fig.~\ref{fig:SBP}). It demonstrates the applicability of the model without necessitating a specific geometry of the magnetic field, when considering a limited and realistic coherence length.
    
    %\item \textbf{Reduced central flux and increased required energy conversion efficiency ($\eta$):} When $l_c$ decreases, the integrated flux around the central regions diminishes. This reduction arises from a greater proportion of particles diffusing further from the source, reducing the localized radiation intensity near the injection point. To compensate for this reduced central flux, a larger conversion efficiency ($\eta$) is necessary to reproduce the observed flux levels in the SBP. This signifies that more energy must be transformed into observable radiation to match the observed profile as coherence lengths shorten.

    \item \textbf{Emergence of a bump-like structure in the SBP:} As we mentioned above, for smaller $l_c$, a significant fraction of particles migrate further into other magnetic coherence at the outer regions. In addition to produce an extended wing-like structure at outer regions, they may also introduce bump-like structures to the SBP. This can be seen, for example, the red curve for Geminga with $l_c=50\,$pc (the top-right panel of Fig.~\ref{fig:SBP} around $\bar{r}=10\,$pc) or the blue curve for Monogem with $l_c=100\,$pc (the middle panel of Fig.~\ref{fig:SBP} around $\bar{r}=25\,$pc). When the distance of the source is large, the bump-like structures would be smeared out by the instrumental PSF, and this is why no such bumps appearing in the case of PSR~J0622+3749 (bottom panels). The reason of the emergence of the bump-like structure can be understood as follows. The direction of the magnetic field in a certain coherence at the outer region may happen to be close to our LOS. This situation becomes more possible when a smaller $l_c$ is considered. As such, due to the projection effect, radiation of electrons propagating along this magnetic coherence add up together in LOS, leading to a bump-like structure in the SBP. Note that the position where the bump appears is not fixed in different realization of simulation, because it depends on the randomly generated magnetic field direction in each coherence. Due to the same reason, the bump-like feature does not always appear in every realization of the simulation. In particularly, when the viewing angle is large, the radiation of electrons from the central magnetic coherence becomes relatively weak, resulting in a more significant contribution from the outer coherence (as illustrated in Appendix Fig.\,\ref{fig:sp_lc}) and the bump-like feature would be more prominent. The formation mechanism of the bump-like feature is similar to the so-called ``mirage'' effect discussed by \citet{Bao24b, YiweiBao24}. 
\end{itemize}

We note that the required viewing angle $\phi$ to match the observed profile also relies on the diffusion parameters. Specifically, larger perpendicular diffusion coefficients ($D_{\perp}$) or larger Alfv{\'e}nic Mach numbers ($M_A$) result in less central concentration of the particle distribution. For a larger $D_\perp$, a smaller $\phi$ is needed to align the model with the observed profile. Finally, besides the 1D SBP, the viewing angle $\phi$ also plays a role in shaping the 2D morphology of the source, particularly in the anisotropy of the morphology. The projected morphology of a pulsar halo on the sky can appear significantly elongated or distorted, depending on the magnetic field geometry and the particle diffusion properties. A detailed analysis of how the halo's morphology relies on the viewing angle will be presented in the subsequent section.

\subsection{Morphology}
To quantitatively evaluate the influence of $l_c$ on the source morphology, we use the Gaussian template and the ellipse Gaussian template, convolved with the PSF, to model the predicted morphology of the source.  The Gaussian template is expressed as:
\begin{eqnarray}
F_{\rm G}(x, y)=F_0\exp \left(-\frac{x^{2}+y^{2}}{2 \sigma_0^{2}}\right)
\end{eqnarray}
where $F_0$ is the normalization of the flux. $(0, 0)$ is the center of the peak of the source. $\sigma_0$ is the standard deviation, assuming it is the same in both $x-$ and $y-$ directions. The ellipse Gaussian template is given by:
\begin{eqnarray}
F_{\rm eG}(x, y)=F_0 \exp \left [  -\frac{1}{2} \left ( \frac{x^2}{\sigma_x^2} + \frac{y^2}{\sigma_y^2} 
 \right ) \right ]
\end{eqnarray}
where $\sigma_x$ and $\sigma_y$ are the standard deviations in the $x-$ and $y-$ directions, respectively. Note that this template has an orientation (if $\sigma_x \neq \sigma_y$) which basically follows the direction of the mean magnetic field of the central magnetic coherence. So we need to rotate the template by an angle $\alpha$ (counterclockwise) to find the best-fit orientation, which can be done via multiplying each point of the template by a rotation matrix $\begin{pmatrix} \sin\alpha & \cos\alpha \\ \cos\alpha & -\sin\alpha \end{pmatrix}$. Therefore, the ellipse Gaussian template has two more free parameters, with one being the orientation of the oval axis, and the other being the second standard deviation. To study the asymmetry of the source, we define $\Delta {\rm TS}_{\rm asy}=2\rm{ln}(\mathcal{L}_{\rm eG}/\mathcal{L}_{\rm G})$, i.e., twice the logarithm of the likelihood ratio of an ellipse Gaussian source assumption to a Gaussian source assumption. $\Delta {\rm TS}_{\rm asy}=29$ corresponds to a significance of 5$\sigma$ for two additional free parameters.

In Appendix Table.~\ref{table1}, we perform the asymmetry test with the 3-year LHAASO full array instrumental response, and present the predicted values of $\rm TS_G$, $\rm TS_{eG}$ and $\rm \Delta TS_{asy}$ based on SBP fittings shown in Fig.~\ref{fig:SBP} for Geminga, Monogem and PSR~J0622+3749. For Geminga and Monogem, LHAASO has not yet published its morphological analyses. Therefore, direct comparisons between our simulations and observational data is not available. Nevertheless, our results show that significant morphological asymmetry of the halos are expected in most of cases considered for the SBP fitting. With future observations of LHAASO, the model can be tested and properties of the interstellar magnetic field can be constrained. It should be also noted that the HAWC collaboration has recently published a new study on the halos of Geminga and Monogem with a larger dateset \cite{2024HAWC}. They divided each of the two halos into four sectors and obtained SBPs of each sector. By fitting the SBP of each sector with the isotropic diffusion model separately, they found the variation among the obtained diffusion coefficients in the four sectors of Geminga reaches of a factor of, which reach a factor of 3.4 for Geminga and a factor of 8 for Monogem. It indicates a noticeable asymmetry in the morphology of the two halos. A more quantitative comparison between our simulations and HAWC’s results could constrain the properties of the interstellar magnetic field around the two pulsars, and will be studied in the future.

For the halo of PSR~J0622+3749, a direct comparison with LHAASO's observation is available. In Fig.~\ref{fig:0622morphology}, we present the simulated $\rm 4^{\circ}\times4^{\circ}$ significance map of the halo of PSR~J0622+3749 at energies above 25\,TeV, under the anisotropic diffusion model with coherence length of 200\,pc, 100\,pc and 50\,pc and $\phi=20^\circ$ (same simulations as that shown in bottom panels of Fig.~\ref{fig:SBP}). The instrumental operation time is set to be 281.9 days and the effective area of LHAASO is taken to be that of the half array, which is the same as in \citet{LHAASO0622}. In the top-right corner of each panel in Fig.~\ref{fig:0622morphology}, we show the TS values obtained using the Gaussian template and the ellipse Gaussian template. Although we can observe the asymmetry of the morphology by eye in the case of $l_c=200\,$pc (the left panel of Fig.~\ref{fig:0622morphology}), the value of $\Delta{\rm TS_{asy}}$ is not large enough to claim the asymmetry. As the coherence length $l_c$ decreases, the values of $\rm TS_{\rm G}$ and $\rm TS_{\rm eG}$ converge, indicating that the ellipse Gaussian template provides less improvement in source detection compared to the Gaussian template. This trend suggests that the source morphology becomes increasingly symmetric with shorter coherence lengths. Indeed, given a smaller coherence length, electrons can cross over a larger number of magnetic coherence around the pulsar and the overall diffusion of particles in the region become more isotropic.

\begin{figure*}[htbp]
\hspace{1.5cm} \includegraphics[width=0.8\textwidth]{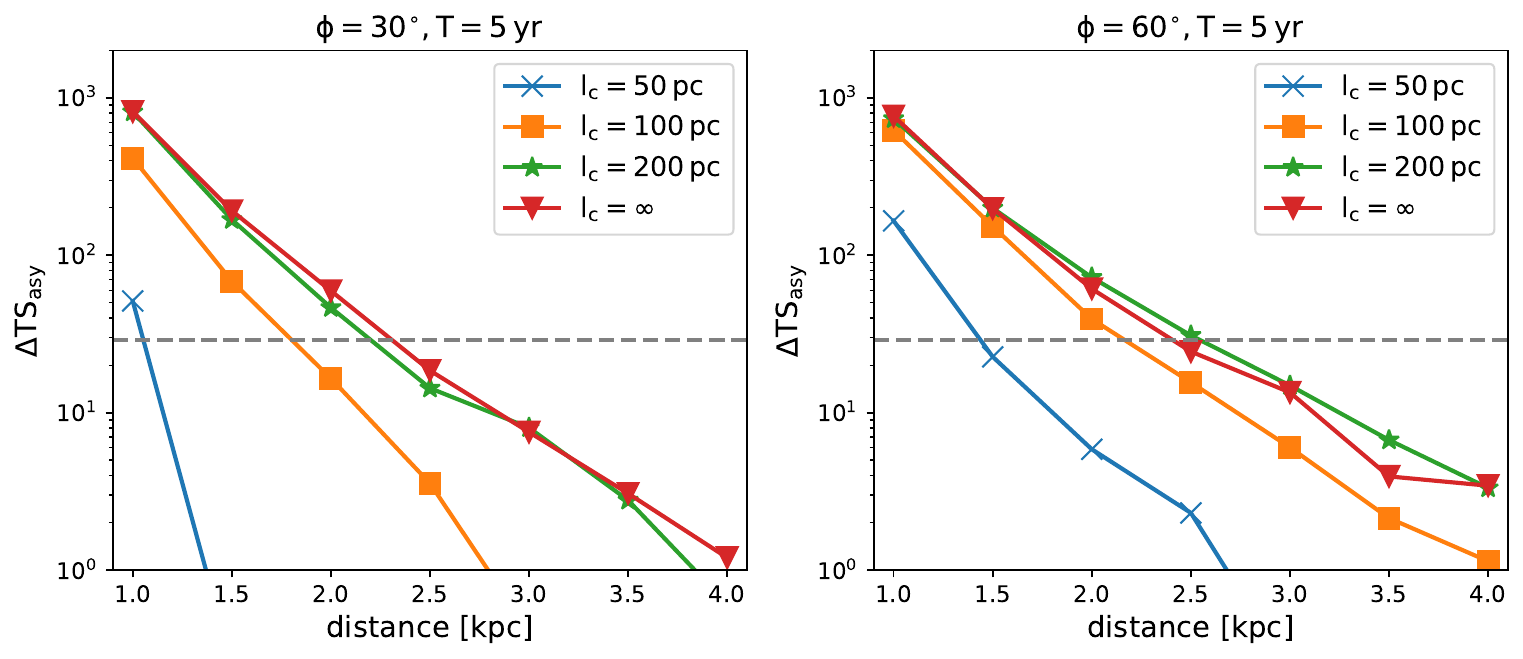}

\caption{Expected significance of asymmetry of sources with different coherence length ($l_c = \infty,200,100,50 \, \rm pc$) at varying distances ($d=1,1.5,2,2.5,3,3.5,4 \, \rm kpc$). The operation time is set to be 5 years. The left(right) panel represents cases with viewing angle of $\phi=30^{\circ}$($\phi=60^{\circ}$). The grey dashed horizontal lines indicate $\Delta TS_{\rm asy}=29$, corresponding to a significance of 5$\sigma$ for asymmetry detection.}
\label{fig:delTS}
\end{figure*}

We further investigate the influence of the source distance on the asymmetry of the morphology. We simulated pulsar halos observed by LHAASO based on the properties of PSR~J0622+3749. We assume the spindown luminosity and age of the pulsar to follow PSR~J0622+3749, and $\eta$ is fixed at 0.5. We perform the simulation under different coherence lengths $l_c$ and different distances $d=\rm 1, 1.5, 2, 3, 3.5, 4\, kpc$. Fig.~\ref{fig:delTS} illustrates the influence of distance and $l_c$ on the source asymmetry, with viewing angles being 30$^{\circ}$ and 60$^{\circ}$. The operation time is set to be 5 years and the effective area of the full array is adopted. As distance increases, the PSF effects become more prominent, reducing the observed source asymmetry. The grey dashed horizontal line in the figure indicates $\Delta TS_{\rm asy}=29$, corresponding a 5$\sigma$ significance level for claiming the source asymmetry. Our results reveal that for large coherence lengths ($l_c \ge \rm 200\,pc$), significant source asymmetry cannot be detected at distances greater than $\sim$2.5 kpc. For relatively small coherence lengths ($l_c \le 50 \, \rm pc$), detecting significant source asymmetry is challenging even at closer distances around 1$\rm \,kpc$. This explains why we have not identified apparent morphological asymmetry so far from detected pulsar halos.

\section{Discussion}

\subsection{Further Smaller Coherence Length $l_c$}
\begin{figure*}[htbp]
\hspace{4cm} 
\includegraphics[width=0.5\textwidth]{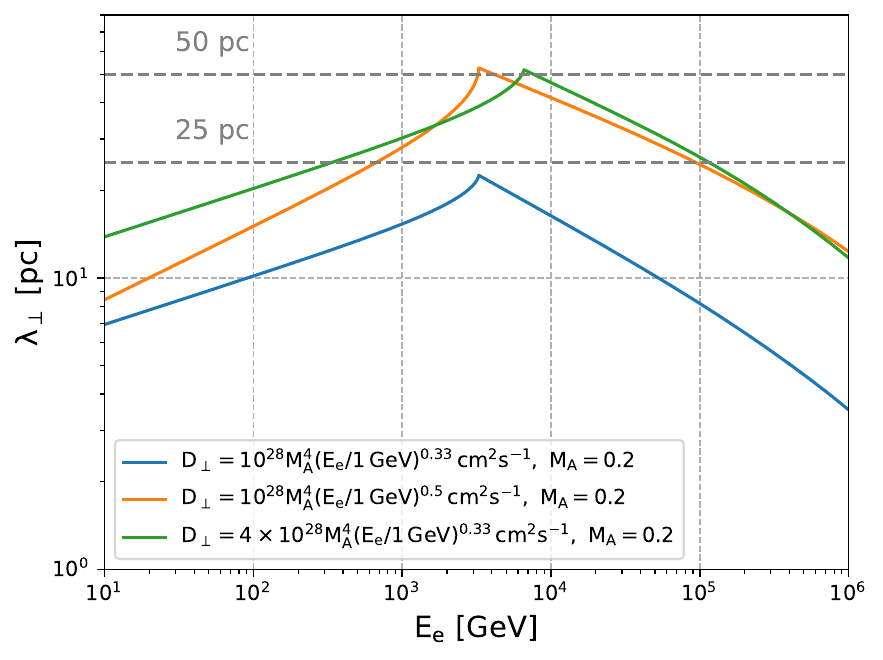}
\caption{Perpendicular diffusion length $\lambda_{\perp}$ of different energies.}
\label{fig:rd}
\end{figure*}

Propagation of particles over a region filled with turbulent magnetic field would show different properties depending on the relation among the magnetic field coherence length $l_c$, the diffusion length of particles and the Larmor radius $r_{\rm L}$ of particles. The typical perpendicular diffusion length $\lambda_\perp$ of these electrons can be estimated by
\begin{eqnarray}
\lambda_{\mathrm{\perp}}=2\left[\int_{E}^{\min \left[E_{0}, E_1\right]} \frac{D_{\perp}(E_{\rm e})}{\dot{E}(E_{\rm e})} \mathrm{d} E_{\rm e}\right]^{1 / 2}
\end{eqnarray}
which represents the maximum length in the direction of perpendicular to the mean magnetic field the injected particles can travel. In Fig.~\ref{fig:rd}, we show the perpendicular diffusion length $\lambda_{\perp}$ at different energies. In this study, we focus on gamma rays with energy above 25\,TeV, which are mainly radiated by electrons with energy above 100\,TeV. $\lambda_\perp$ of these electrons are generally smaller than 25\,pc. On the other hand, we assume the mean magnetic field direction keeps the same within a cube of size $l_c$. For the considered magnetic field coherence length $l_{\rm c}\geq 50\,$pc, most of these electrons will be confined within the coherent magnetic region and do not escape via perpendicular diffusion, i.e., $\lambda_\perp <l_c$. If the coherence length is significantly smaller than $\lambda_\perp$ (i.e., $l_c \sim 10\,$pc), a considerable fraction of particles may escape from the tube, and our present method cannot deal with the further propagation of these escaped particles. More sophisticated treatment such as test-particle simulation with a more realistic magnetic field configurations may give a more reasonable description of the particle propagation in this scenario.

For further smaller coherence length ($\l_c \lesssim 1\, \rm pc$), \citet{Lopez18} have shown that the rapid spatial variation of the mean field direction can confine particles, resulting in a small diffusion coefficient. Because particles traverse a large number of coherent magnetic region, the overall diffusion becomes isotropic. This scenario provides another possible explanation of the observation of pulsar halos. We do not delve into this scenario in this study, which falls outside the framework of the anisotropic diffusion model.

%In this studies, we mainly explored scenarios where the coherence length $l_{\rm c}$ of the magnetic field is greater than or equal to 50\,pc. Under these conditions, the perpendicular diffusion length $\lambda_{\perp}$ is smaller than 25\,pc, as shown in Fig.~\ref{fig:rd}. Given that the injection point is set at the center of one magnetic field coherence, this setup ensures that most particles are well confined in the direction perpendicular to the mean magnetic field.

%When the coherence length decreases to values less than 50 pc, the situation becomes significantly more complex. Some particles may escape from the magnetic field tube, necessitating the consideration of more realistic and intricate magnetic field configurations. For even smaller coherence lengths ($\l_c \lesssim 1\, \rm pc$) with a strong mean magnetic field, particles are confined by the highly curved yet strong magnetic field. This confinement results in a small diffusion coefficient, though isotropic on large scales. This scenario represents another mechanism that suppresses the diffusion coefficient, which falls outside the scope of the anisotropic diffusion model \citep{Lopez18}. Consequently, we do not delve into these cases in this study.

Finally, for very small coherence lengths which are even smaller than particle's Larmor radius, i.e., $l_{\rm c} \lesssim r_{\rm L}$, particles cannot be effectively scattered in each coherence length. Within each coherence length, quasi-rectilinear propagation of particles occurs. However, this scenario is unlikely to appear in pulsar halos, where the typical Larmor radius of particles reads
$r_L \sim 0.02 (E_e/100\, {\rm TeV})(B/5\,{\mu \rm G}) \,$pc.
%but could exist in the intracluster medium for ultrahigh-energy cosmic rays. %Magnetic field in this environment can be amplified by the fluctuation dynamo on scales smaller than the turbulence injection scale \citep{Seta21prf, ChaoZhang24}, and a detailed numerical simulation of MHD turbulence to be fully understood. Such simulations would account for the intricate interactions between cosmic rays and the highly turbulent magnetic field structures, providing deeper insights into the behavior of particles under these conditions.
 
\subsection{Influence of Other Model Parameters}
The main purpose of this paper is to explore the effects of coherence length $l_c$ on the observational characteristics of pulsar halos, and hence we fix some model parameters for simplicity. However, some model parameters can also influence the predicted TS value of simulated pulsar halos. In this section, we will briefly discuss the impacts of these parameters.

\paragraph{Diffusion Coefficient}
The benchmark value for the diffusion coefficient used in this study is $D_{\parallel}=10^{28}(E_e / 1 \, \rm{GeV})^{1/3} \, \rm cm^2s^{-1}$, with $D_{\perp}=D_{\parallel}M_A^4$ and $M_A=0.2$. A larger diffusion coefficient would result in a flatter SBP. To fit the observed SBP, a smaller viewing angle $\phi$ is required. On the other hand, increased diffusion coefficients lead to a more extended morphology. This extended morphology dilutes the source intensity, reducing the TS value and making the source harder to be detected and resolved with current instruments, unless the pair conversion efficiency $\eta$ is larger. 
%In addition, as we discussed in the previous section, if larger diffusion coefficient is employed, the perpendicular diffusion length is also larger and our current method  coefficients, the maximum diffusion length would exceed the radius of the magnetic tube of $l_c=50\, \rm pc$, which also serves as the calculation boundary for magnetic field configuration. In such cases for $l_c=50\, \rm pc$, a more realistic and complex magnetic field configuration must be considered. This includes accounting for the transition of particle propagation from regions dominated by regular magnetic field to those dominated by chaotic magnetic field. For $l_c\ge100\, \rm pc$, this magnetic field configuration remains applicable, ensuring that particles remain confined and follow predictable diffusion patterns.

\paragraph{Alf{\'e}nic Mach number $M_A$} 
In this study, we adopt a sub-Alf{\'e}nic regime with the Alf{\'e}nic Mach number fixed at $M_A=0.2$. This condition ensures a highly anisotropic intrinsic particle distribution, with a strong preference for the direction parallel to the mean magnetic field compared to the perpendicular direction. For a larger value of $M_A$, the intrinsic particle distribution would become less anisotropic. Consequently, the influence from the viewing angle $\phi$ and coherence length $l_c$ on the expected morphology would be less pronounced. Additionally, a larger $M_A$ would also lead to a more extended particle distribution and reduce the expected TS value of the sources, making it more challenging to detect and resolve the sources \citep{KaiYan22apj}. In the super-$\rm Alf\acute{e}nic$ regime, the magnetic fields are highly turbulent and disordered. Particles diffusion will not follow a certain preferential direction and is subject to the local magnetic field structure. A representative study of this regime is presented by \citet{Lopez18} as mentioned above, using the test-particle simulation with synthetic turbulence. Note that the properties of turbulence may vary from place to place in the ISM, so the models built under the two scenarios (i.e., sub-Alfv{\'e}nic and super-Alfv{\'e}nic) are complementary instead of mutually exclusive. Either of them could be applicable to a pulsar halo, depending on the environment around the pulsar.

\paragraph{Injection Spectrum} 
\citet{2019Xi} reported a null detection of the diffuse multi-GeV emission from the vicinity of Geminga by Fermi-LAT, suggesting a hard injection spectrum of electron from the PWN. \citet{LHAASO0622} also employed a hard injection spectrum with $p < 2$ to model the halos of PSR~J0622+3749,  in order to reconcile with non-detection of the halo's GeV counterpart. Following these observations, the spectral index of the injection electrons $p$ are set at 1.6. The injection spectral index is not a crucial parameter  this study, because it can merely affect the source morphology. From the perspective of energy budget, on the other hand, a softer/harder injection spectrum would lead to a lower/higher flux above 25\,TeV. To fit a certain observed flux, a higher/lower energy conversion efficiency $\eta$ is then required. This parameter can be better constrained when spectra of pulsar halos are measured more accurately.

\section{Conclusion}
Particles in pulsar halo may traverse multiple magnetic coherence. In this study, we performed a dedicated study of the morphology of TeV pulsar halos under the anisotropic diffusion model with taking into account a limited coherence length $l_c=50-200$\,pc for the interstellar magnetic field. To model particle diffusion in this scenario, we developed a random walk simulation that accounts for the stochastic nature of particle transport across multiple magnetic coherence lengths. Besides, we incorporate the instrument response by considering the effects of the background noise from the cosmic ray, the PSF and made quantitative simulations of 2D TS maps of pulsar halos expected in LHAASO's measurements. We explored the surface brightness profile and the asymmetry of pulsar halos expected under the anisotropic diffusion model with a limited coherence length of magnetic field $l_c$. 
Our main findings are summarized as follows:
\begin{itemize}
    \item The observed SBPs of pulsar halos can be well reproduced with a much broader range of the viewing angle $\phi$, which largely relaxes the restriction of $\phi\leq 5^\circ$ as required in the case of an infinite $l_c$. 
    \item The asymmetry of the source morphology is less significant, because particles traversing multiple coherence leads to a more isotropic particle distribution over the entire halo. The smaller $l_c$ is, the more prominent this effect is.
\end{itemize}
We also investigated the influence of the source distance on the the source asymmetry. As the distance increases, the PSF of the instrument becomes increasingly significant. It further smooths out the asymmetry in the source morphology. As a result, when fitting the source of intrinsic asymmetry with an ellipse Gaussian template, it would not significantly improve the TS value with respect to the fitting with a symmetric Gaussian template. Our study provides an explanation of why we have not observed apparent asymmetric morphology in pulsar halos so far. Nevertheless, with a longer exposure time, the statistics would be eventually high enough to identify the asymmetry in the morphology of pulsar halos and their candidates (if there is intrinsic asymmetry). On the other hand, observations of instruments with a smaller PSF can also help in identifying the asymmetry. In the future, longer observations of LHAASO and HAWC, together with high-resolution image provided imaging air Cherenkov telescopes such as CTA \citep{CTA2011}, ASTRI \citep{ASTRI-MINI22} and LACT \citep{LACT2024} will provide valuable insights into the observational characteristics of pulsar halos. These studies will facilitate our understanding of cosmic-ray propagation in the interstellar medium and finally help to unravel the puzzle of the origin of cosmic rays.

\section*{Acknowledgements}
{We thank Hai-Ming Zhang for the helpful discussions. This work is supported by the National Science Foundation of China under grants No.12393852 and No.12333006.

\appendix

\section{Random Walk Simulation}
Consider a random walker on a three-dimensional lattice that hops to eight directions $\rm (\pm \delta_x,\pm \delta_y,\pm \delta_z)$ with equal probabilities 1/8 in a single step. Let $\rm N(x,y,z,t)$ be the number density that the particle is at site $\rm (x, y, z)$ at the t time step. Then evolution of this number density is described by the equation
\begin{eqnarray}
N(x, y, z, t+\tau)=\frac{1}{8} N(x-\delta_x, y-\delta_y, z-\delta_z, t)+\frac{1}{8} N(x+\delta_x, y-\delta_y, z-\delta_z, t) \\
+\frac{1}{8} N(x-\delta_x, y+\delta_y, z-\delta_z, t)+\frac{1}{8} N(x-\delta_x, y-\delta_y, z+\delta_z, t) \\
+\frac{1}{8} N(x+\delta_x, y+\delta_y, z-\delta_z, t)+\frac{1}{8} N(x+\delta_x, y-\delta_y, z+\delta_z, t) \\
+\frac{1}{8} N(x-\delta_x, y+\delta_y, z+\delta_z, t)+\frac{1}{8} N(x+\delta_x, y+\delta_y, z+\delta_z, t)
\label{eq:P_N}
\end{eqnarray}
The diffusion equation can be obtained by the Taylor expansion of the above formula which is reserved to a small quantity of the second order in time and to a small quantity of the third order in space 
\begin{eqnarray}
\frac{\partial N}{\partial t}=\frac{\delta_x^{2}}{2 \tau} N_{\rm x x}(x, y, z, t)+\frac{\delta_y^{2}}{2 \tau} N_{\rm y y}(x, y, z, t)+\frac{\delta_z^{2}}{2 \tau} N_{\rm z z}(x, y, z, t)+o\left(\delta^{3}\right)+o\left(\tau^{2}\right)
\label{eq:N_diffusion}
\end{eqnarray}
Substituting $\delta_x^{2}/2\tau$ to $D_x$,  $\delta_y^{2}/2\tau$ to $D_y$ and  $\delta_z^{2}/2\tau$ to $D_z$, we get $N_{\rm t}(x, y, z, t) = \nabla \cdot[\mathbb{D} \cdot \nabla N(x,y,z, t )]$, where $\mathbb{D}$ is a diagonal matrix with diagonal elements being $\rm D_x,D_y,D_z$.

Starting from the evolution equation of random walk particles, we derived the diffusion equation. This means that the distribution of particles undergoing a random walk inherently provides the solution to the diffusion equation. By modeling the random walk, where particles move in a series of random, independent steps characterized by a probability distribution, we can describe how particles spread over time. This approach not only simplifies the simulation of complex diffusion processes but also ensures that the particle distribution aligns with the theoretical expectations of the diffusion equation.

\section{Comparison with analytical solutions}
The analytical expression for the particle distribution $H(\vec{r},E_e)$ has been explored under some specific cases.

For isotropic or two-zone diffusion, in which the diffusion coefficient is equal to $D_0$ in the near source region within a certain radius $r_0$ and equal to $D_1$ beyond $r_0$. The analytical was given by \citet{Schroer23prd}
\begin{eqnarray}
\mathcal{H}(r, E_e, t)=\int_{0}^{\infty} \mathrm{d} \psi \frac{\xi e^{-\psi}}{\pi^{2} \lambda_{0}^{2}\left[A^{2}(\psi)+B^{2}(\psi)\right]}\left\{\begin{array}{ll}
r^{-1} \sin \left(2 \sqrt{\psi} \frac{r}{\lambda_{0}}\right) & 0<r<r_{0} \\
A(\psi) r^{-1} \sin \left(2 \sqrt{\psi} \frac{r \xi}{\lambda_{0}}\right)+B(\psi) r^{-1} \cos \left(2 \sqrt{\psi} \frac{r \xi}{\lambda_{0}}\right) & r \geq r_{0}
\end{array}\right.
\end{eqnarray}
where 
\begin{eqnarray}
\begin{array}{c}
A(\psi)=\xi \cos (\chi) \cos (\xi \chi)+\sin (\chi) \sin (\xi \chi)+\frac{1}{\chi}\left(\frac{1-\xi^{2}}{\xi} \sin (\chi) \cos (\xi \chi)\right), \\
B(\psi)=\frac{\sin (\chi)-A(\psi) \sin (\xi \chi)}{\cos (\xi \chi)} .
\end{array}
\end{eqnarray}
The quantity $\chi=2\sqrt{\psi}\frac{r_0}{r}$, $\xi=\sqrt{D_0/D_1}$ and $\lambda_0=\left[\int_{E_{\rm e}}^{E_0} \frac{D_0(E_{\rm e})}{\dot{E}(E_{\rm e})} \mathrm{d} E_{\rm e}\right]^{1 / 2}$.  

For anisotropic diffusion with $l_c=\infty$, the analytical expression can be obtained as
\begin{eqnarray}
\mathcal{H}(r, z, E_e, t) = \frac{1}{\left(\pi \right)^{3 / 2} \lambda_{\perp}^{2} \lambda_{\parallel}} \exp \left(-\frac{r^{2}}{\lambda_{\perp}^{2}}\right)\exp \left(-\frac{ z^{2}}{\lambda_{\parallel}^{2}}\right)
\end{eqnarray}
%\\
%= \frac{M_A^2}{\left(\pi \lambda_{\perp}^{2}\right)^{3 / 2}} \exp \left(-\frac{r^{2}}{\lambda_{\perp}^{2}}\right)\exp \left(-\frac{ M_A^4 z^{2}}{\lambda_{\perp}^{2}}\right)  \\
%= \frac{1}{\left(\pi \lambda_{\parallel}^{2}\right)^{3 / 2}M_{\rm A}^4} \exp \left(-\frac{r^{2}}{\lambda_{\parallel}^{2}M_{\rm A}^4}\right)\exp \left(-\frac{ z^{2}}{\lambda_{\parallel}^{2}}\right) 

In Fig.~\ref{fig:Hr}, we show the comparisons of particle distribution from random walk simulation with analytical solutions. We see that they are in good consistency.

\begin{figure*}[htbp]
\hspace{1.5cm} \includegraphics[width=0.8\textwidth]{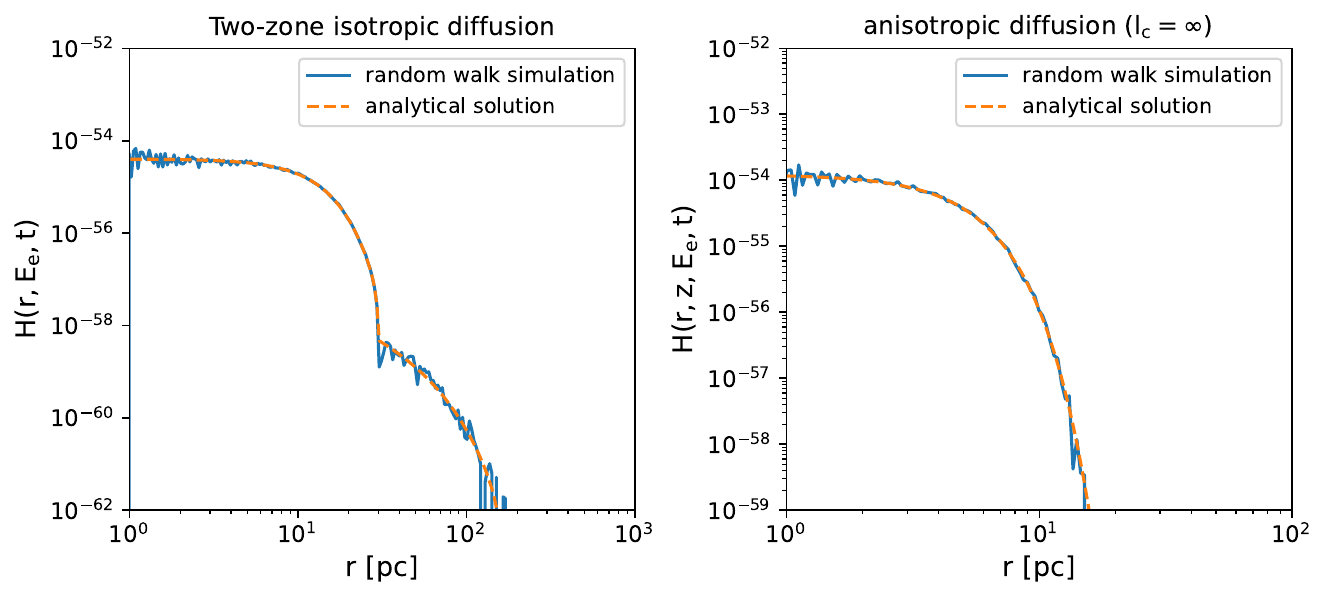}
\caption{Comparison of particle distribution from random walk simulation (blue lines) with analytical solutions (orange dashed lines). The particle energy is set to be $1 \, \rm TeV$. The injection time is set to be 10 kyr. For two-zone isotropic model, $D_0=10^{26}(E_e / 1 \, \rm{GeV})^{0.33} \, \rm cm^2s^{-1}$, $D_1=10^{28}(E_e / 1 \, \rm{GeV})^{0.33} \, \rm cm^2s^{-1}$, $r_b=30 \, \rm pc$. In the anisotropic diffusion model, $D_{\parallel}=10^{28}(E_e / 1 \, \rm{GeV})^{0.33} \, \rm cm^2s^{-1}$, $M_A=0.2$. In the right panel, we set $z= 20\, \rm pc$ for illustration purposes.}
\label{fig:Hr}
\end{figure*}

\section{Dependence of 1D gamma-ray SBPs on Parameters}
In Fig.~\ref{fig:sp_phi} and Fig.~\ref{fig:sp_lc}, we show how the expected 1D gamma-ray SBPs obtained with our random walk simulation change with parameters of magnetic fields. More specifically, Fig.~\ref{fig:sp_phi} compares the results for $l_{\rm c}=50\,pc$, 100\,pc, 200\,pc and $\infty$ under a given viewing angle. Fig.~\ref{fig:sp_lc} exhibits the dependence of SBP on $\phi$ from $\phi=0^\circ$ to $\phi=60^\circ$ with an increment of $5^\circ$ under a given coherence length. 

\begin{figure*}[htbp]
\includegraphics[width=1\textwidth]{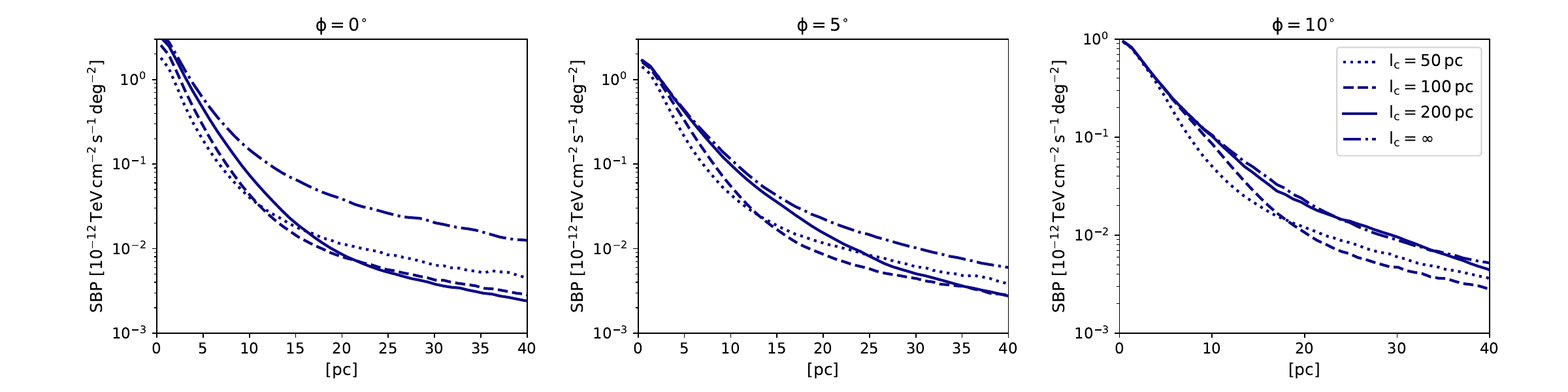}

\caption{SBPs for coherence lengths ($l_c = \infty, 200, 100, 50\ \mathrm{pc}$) at viewing angles ($\phi = 0^\circ, 5^\circ, 10^\circ$). As $l_c$ decreases, outer-region radiation bends outward, limiting observable emission to regions near the injection point. The sharper particle density and radiation gradients near the injection point result in a steeper SBP.}
\label{fig:sp_phi}
\end{figure*}

\begin{figure*}[htbp]
\hspace{0.5cm} \includegraphics[width=1\textwidth]{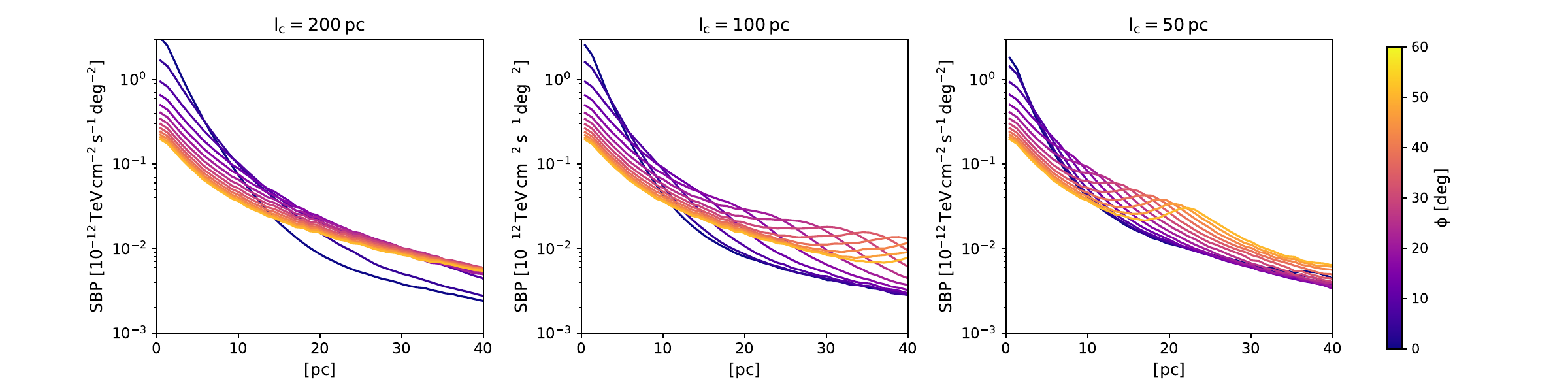}
\caption{SBPs under different viewing angles (lines shown at $5^{\circ}$ intervals) for coherence lengths $l_c = 200, 100, 50\ \mathrm{pc}$. When multiple coherence lengths are considered, bumps begin to appear in the SBP for large $\phi$. The smaller the $l_c$, the earlier and more prominent these bumps become, highlighting the significant contribution of outer-region radiation.}
\label{fig:sp_lc}
\end{figure*}

\section{Prediction of LHAASO observations on pulsar halos with different templates}
 In Tab.~\ref{table1}, we show the expected TS values obtain with the mock LHAASO data of the pulsar halos, analyzed by the 2D symmetric Gaussian template and the ellipse Gaussian template. The mock data is generated by two steps. First, we simulate the 2D intensity maps of halos of Geminga, Monogem and PSR~J0622+3749 with our method. For each pulsar halo, we generate 12 intensity maps under different combinations between $\phi$ and $l_{\rm c}$, same as those shown in Fig.~\ref{fig:SBP}. We then simulate the LHAASO observations of these halos, employing the 3-year LHAASO full-array instrumental response. Finally, we analyze the mock LHAASO data with the symmetric and asymmetric Gaussian templates and obtain the TS values. 

%\centering
\begin{table}[htbp]
\centering
\caption{ Comparison of $\rm TS_G$, $\rm TS_{eG}$ and $\rm \Delta TS_{asy}$ under the parameters of SBP fitting in Fig.~\ref{fig:SBP} for Geminga, Monogem and PSR J0622+3749. The instrument response is the based on LHAASO with an operation time of 3 years.}
\begin{tabular}{cc|ccc|ccc|ccc}
\hline
\hline
\multicolumn{1}{c}{Source Name} 
& \multicolumn{1}{c|}{$\rm \phi$} 
& \multicolumn{3}{c|}{$l_c = 200~\mathrm{pc}$} 
& \multicolumn{3}{c|}{$l_c = 100~\mathrm{pc}$} 
& \multicolumn{3}{c}{$l_c = 50~\mathrm{pc}$} \\
 & & $\rm TS_G$ & $\rm TS_{eG}$ & $\rm \Delta TS_{asy}$ 
 & $\rm TS_G$ & $\rm TS_{eG}$ & $\rm \Delta TS_{asy}$ 
 & $\rm TS_G$ & $\rm TS_{eG}$ & $\rm \Delta TS_{asy}$ \\
\hline
Geminga & $\rm 10^{\circ}$ & 6895.43 & 9114.57 & 2219.14 & 6676.05 & 7633.53 & 957.48 & 6174.06 & 6219.18 & 45.12 \\
Geminga & $\rm 20^{\circ}$ & 7738.5 & 9274.79 & 1536.28 & 7897.25 & 12481.81 & 4584.55 & 8185.5 & 9098.57 & 913.07 \\
Geminga & $\rm 30^{\circ}$ & 7668.79 & 15941.66 & 8272.87 & 7606.24 & 15145.63 & 7539.39 & 8124.72 & 10264.48 & 2139.76 \\
Geminga & $\rm 40^{\circ}$ & 7802.79 & 17686.83 & 9884.03 & 7532.78 & 16767.86 & 9235.08 & 8088.61 & 11682.88 & 3594.27 \\
Monogem & $\rm 10^{\circ}$ & 1334.25 & 1773.85 & 439.6 & 1327.27 & 1522.68 & 195.41 & 1086.8 & 1093.5 & 6.7 \\
Monogem & $\rm 20^{\circ}$ & 1359.26 & 2616.33 & 1257.07 & 1486.42 & 2358.89 & 872.47 & 1620.38 & 1833.37 & 212.98 \\
Monogem & $\rm 30^{\circ}$ & 1472.06 & 3355.49 & 1883.42 & 1521.84 & 3159.82 & 1637.97 & 1710.44 & 2201.86 & 491.42 \\
Monogem & $\rm 40^{\circ}$ & 1442.13 & 3400.58 & 1958.45 & 1276.37 & 3065.37 & 1789.0 & 1701.29 & 2468.72 & 767.43 \\
PSR J0622+3749 & $\rm 10^{\circ}$ & 999.41 & 1014.53 & 15.12 & 906.41 & 906.27 & 0.14 & 865.28 & 880.4 & 15.11 \\
PSR J0622+3749 & $\rm 20^{\circ}$ & 1147.16 & 1301.76 & 154.6 & 1163.51 & 1182.92 & 19.41 & 1042.69 & 1052.13 & 9.44 \\
PSR J0622+3749 & $\rm 30^{\circ}$ & 1063.46 & 1388.2 & 324.74 & 1078.21 & 1177.02 & 98.81 & 1136.15 & 1136.33 & 0.19 \\
PSR J0622+3749 & $\rm 40^{\circ}$ & 1026.13 & 1486.92 & 460.79 & 1061.88 & 1255.38 & 193.51 & 1178.66 & 1184.95 & 6.29 \\

\hline
\hline
\end{tabular}

\label{table1}
\end{table}

%\bibliography{sample631}
%\bibliographystyle{aasjournal}

\end{document}